\documentstyle[epsfig,aps,preprint]{revtex}

\begin{document}
\draft
\title{ CLASSICAL AND QUANTUM INTERACTION OF THE DIPOLE}
\author{J. Anandan}
\address{ Department of Physics and Astronomy, University of
South
Carolina,\\ Columbia, SC 29208. \\E-mail: jeeva@sc.edu}
\date{Oct. 4, 99}
\maketitle

\begin{abstract}
A unified and fully relativistic treatment of the
interaction of the
electric and magnetic dipole moments of a particle with the
electromagnetic
field is given. New forces
on the particle due to the combined effect of electric and
magnetic dipoles are
obtained. Four new
experiments are proposed, three of which would observe
topological phase shifts.
\end{abstract}
\bigskip

A beautiful aspect of electromagnetism, which is becoming
increasingly
fundamental, is the duality between
electric and magnetic fields. It is therefore
natural to extend the many interesting facets of the
influence of the
electromagnetic field on the magnetic dipole [1-7] to the
dual situations \cite{wi1994,sp1998}.
In this letter, I shall give a model independent and fully
relativistic
derivation of the low
energy Lagrangian for the interaction of the
electric and
magnetic dipoles with an external
electromagnetic field \cite{pu1950}.
This treatment applies to elementary particles, nuclei, atoms and
molecules that have an electric or magnetic dipole or both. New forces
on the dipole
will be obtained using
the non abelian nature of the interaction
\cite{go1989,an1989,an1989a}. I shall
then propose
new feasible experiments to test this interaction,
three of which would
have topological aspects analogous
to the Aharonov-Bohm (AB) effect \cite{ah1959}.

The simplest classical Lorentz invariant action for a
particle of
mass $m$ interacting with the electromagnetic field strength
$F_{\mu\nu}$ is
\begin{equation}
I=-\int{mc ds} -{1\over 2c} \int{F_{\mu\nu}D^{\mu\nu} ds}
\label{action}
\end{equation}
where $s$ is the proper time along the particle's world-
line. For simplicity
the total electric and magnetic charges of the particle
have been assumed to be zero. $ D^{\mu\nu}$ may be taken
to be anti-symmetric
without loss of generality, and will be called the dipole
moment tensor. More
generally, $I=\int {\cal L} ds$, where $\cal L$ is a Lorentz
scalar that is an analytic function of $F_{\mu\nu}$. In the
power series
expansion of $\cal L$ in $F_{\mu\nu}$, the sum of the first
and higher order
terms is the last term in (\ref{action}). In the
corresponding power
series for $ D^{\mu\nu}$, the term that is independent of
$F_{\mu\nu}$ is called
the intrinsic dipole moment and the remainder is called the
induced dipole
moment.

In any frame, $D^{0i}$ and $D^{ij}$, that couple,
respectively,
to the electric field components $F_{0i}$ and the magnetic
field components
$F_{ij}$, are called the components of the electric and
magnetic dipole moments. In the
rest frame of the particle, $D^{0i}$ and $D^{ij}$ represent
the intrinsic
electric and magnetic dipole moments, much like the energy
in the rest frame
representing the mass of the particle. These may be
covariantly represented, respectively, by the electric and
magnetic dipole
moment 4-vectors
\begin{equation}
d^\mu \equiv {D^\mu}_\nu v^\nu , ~~m^\mu\equiv {1\over 2}
\epsilon^{\mu\nu\rho\sigma} D_{\nu\rho}v_\sigma
\end{equation}
where the 4-velocity $ v^\mu={dx^\mu\over ds}$. Then
\begin{equation}
d_\mu v^\mu =0, ~~ m_\mu v^\mu =0.
\end{equation}
This reflects the fact that only
three components of each dipole
moment 4-vector
are independent, because in the rest frame,
in which $v^\mu=(1,0,0,0)$, each vector has
only three
(spatial) components.

Now the identity
$ \epsilon^{\mu\nu\rho\sigma}
\epsilon_{\mu\alpha\beta\gamma}=-6
\delta^\nu_{[\alpha}\delta^\rho_\beta\delta^\sigma_{\gamma]}
$,
implies \cite{dual}
\begin{equation}
{1\over 2} F_{\mu\nu}D^{\mu\nu}=d^\mu F_{\mu\nu} v^\nu +
m^\mu {F*}_{\mu\nu}
v^\nu
\label{decomposition}
\end{equation}
The electric and magnetic fields in the rest frame may also
be covariantly
defined by
\begin{equation}
E^\mu = {F^\mu}_\nu v^\nu , ~~B^\mu= {1\over 2}
\epsilon^{\mu\nu\rho\sigma}
F_{\nu\rho}v_\sigma={{F*}^\mu}_\nu v^\nu
\label{EB}
\end{equation}
Then $ E^\mu v_\mu = 0 = B^\mu v_\mu$. In the rest frame, $
E^\mu =(0, {\bf E}
)$ and $ B^\mu =(0,{\bf B})$,
where $\bf E$ and $\bf B$ are the electric and magnetic
(spatial) vectors, in the usual notation.
On using (\ref{EB}), (\ref{decomposition}) reads
\begin{equation}
{1\over 2} F_{\mu\nu}D^{\mu\nu}=d^\mu E_\mu + m^\mu B_\mu
\label{restframe}
\end{equation}
On defining the 4-vector potentials,
\begin{equation}
D_\nu = d^\mu F_{\mu\nu}, ~~M_\nu= m^\mu {F*}_{\mu\nu}
\label{dipoles}
\end{equation}
(\ref{decomposition}) may also be rewritten as
\begin{equation}
{1\over 2} F_{\mu\nu}D^{\mu\nu}=D_\mu v^\mu + M_\mu
v^\mu=a_\mu v^\mu
\label{interaction}
\end{equation}
where $a_\mu =D_\mu + M_\mu$.

From (\ref{action}) and (\ref{interaction}),
the relativistic lagrangian
is
\begin{equation}
L_R=-mc {ds\over dt} - M_\mu {dx^\mu\over dt} - D_\mu
{dx^\mu\over dt}
\label{lagrangian}
\end{equation}
Clearly then, the duality between the interactions of the
electric and magnetic
{\it dipoles} is complete to all orders of $v/c$
indepenently whether or not
there are magnetic
monopoles.
Each of the last two terms in (\ref{lagrangian})is analogous
to the
electromagnetic interaction term
$eA_\mu {dx^\mu\over dt} $ in the Lagrangian for a charged
particle, where $A_\mu$ is the electromagnetic
4-vector potential. This
immediately suggests
topological effects analogous to the AB effect
\cite{ah1959}.

When the Lagrangian is quantized, the phase shift that the
particle experiences
due to the the electromagnetic field, using the action
(\ref{action}), and
(\ref{interaction}), is given by the phase factor
\begin{equation}
P\exp (-{i\over 2c\hbar} \int_\gamma F_{\mu\nu}D^{\mu\nu}
ds)
= P\exp (-{i\over \hbar c} \int_\gamma a_\mu dx^\mu)
\label{phase}
\end{equation}
where P represents path ordering because $D^{\mu\nu}, D_\mu,
M_\mu$
and therefore $a_\mu$ are now operators which need not
commute. For a
topological phase, the effect of
(\ref{phase}) on the wave function should not change when
$\gamma$ is deformed
in a suitable region. Such a deformation may be obtained by
infinitesimal
deformations. To see the effect of an infinitesimal
deformation, consider
(\ref{phase}) around an infinitesimal closed curve spanning
a surface element
$d\sigma^{\mu\nu}$:
\begin{equation}
P\exp (-{i\over \hbar c} \oint_\gamma a_\mu dx^\mu)=
1-{i\over 2\hbar c} G_{\mu\nu} d\sigma^{\mu \nu}
\label{infinitesimal}
\end{equation}
where the ``Yang-Mills field strength'' $G_{\mu\nu}$ is
defined by
\begin{equation}
G_{\mu\nu} = \partial_\mu a_\nu - \partial_\nu a_\mu
-{i\over \hbar c}[a_\mu , a_\nu]
\end{equation}

An external field modifies the motion of a wave function,
via Huyghen's
principle, by the phase shift that it causes in the
interference of secondary
wavelets. This gives a relativistic correspondence principle
that enables the
classical equation of motion to be determined from the phase
shift around an
infinitesimal closed curve \cite{an1977}. Let $ \xi^\mu
(t)=(t,
<\psi(t)|x^i|\psi(t)>)$ represent the world-line of a
wave-packet
$ |\psi(t)> $ of the particle, where $x^i$ are the
position operators. Then for a suitably localized
wave-packet, the 4-force
\begin{equation}
f^\mu\equiv m {d^2\xi^\mu\over ds^2}=
<\psi|{G^{\mu}}_\nu|\psi>{d\xi^\mu\over
ds}
\label{force}
\end{equation}
to a good approximation.
The approximation to (\ref{force}) in the low energy
limit, to $O(1/c)$, was obtained
by writing $d<x^i>/dt$ and $d^2<x^i>/dt^2$ in terms of the
commutators of the
Hamiltonian \cite{an1989}. Then Lorentz covariance implies
that (\ref{force}) is valid to all orders in $1/c$.
A purely classical
derivation of
(\ref{force}), with the expectation values replaced by
the corresponding classical quantities, may
be obtained by replacing the above quantum commutators by
appropriate Poisson
brackets of the relativistic classical Hamiltonian.
The 4-force (\ref{force}) is also
analogous to the 4-force on a Yang-Mills particle obtained in
the classical limit
of the Dirac's equation by Wong \cite{wo1970}.

(\ref{infinitesimal}) and (\ref{force}) show that
$G_{\mu\nu}$ determines {\it
both} whether the phase shift is topological (i.e. whether
(\ref{phase}) is
unchanged when $\gamma$ is varied) and non local (the force
on the interfering beams is zero). If
\begin{equation}
G_{\mu\nu} =0
\label{strong}
\end{equation}
everywhere in a non simply connected region to which the
interfering beams are
confined, then the phase shift will be called {\it strongly
topological or non local}.
The phase shift will be called {\it weakly topological or non
Local} if
\begin{equation}
<\psi|G_{\mu\nu}|\psi> =0
\label{weak}
\end{equation}

It is interesting that the last term in (\ref{action}),
which appears to be a non gauge interaction,
may be regarded as a gauge
interaction in the
fully relativistic theory because of the identity
(\ref{interaction}). Hence,
the topological
and geometrical phases implied by the Lagrangian
(\ref{lagrangian}) are also
valid to all orders in $1/c$, unlike in earlier treatments.
I shall now make
a low energy approximation by neglecting terms of
$O(1/c^2)$. From
(\ref{dipoles}) and (\ref{lagrangian}),
the parts of $D_\mu$ and $M_\mu$ that
contribute to $ L_R$ in
this approximation are
\begin{eqnarray}
D_\mu &=& (-{\bf d}\cdot {\bf E}, {\bf d}\times {\bf B}), M_\mu
=(-{\bf\mu}\cdot
{\bf B}, -{\bf\mu}\times {\bf E}),
\nonumber \\&& a_\mu=M_\mu +D_\mu
\label{ledipoles}
\end{eqnarray}
The Lagrangian obtained in this approximation from
(\ref{lagrangian}),
after subtracting the rest mass energy, is
\cite{previous}
\begin{equation}
L ={1\over 2}mv^2 +{\bf\mu}\cdot {\bf B} +
{\bf v}\cdot {\bf\mu}\times {\bf E}+
{\bf d}\cdot {\bf E}- {\bf v}\cdot {\bf d}\times {\bf B}
\end{equation}
Then the Hamiltonian, obtained by the usual
Legendre transformation from this Lagrangian, is
\begin{eqnarray}
H &=& {1\over 2m} ({\bf p}-{\bf\mu}\times {\bf E}+{\bf d}\times
{\bf B})^2 -
{\bf\mu}\cdot {\bf B}-{\bf d}\cdot {\bf E}
\nonumber \\&& \equiv {1\over 2m} ({\bf p}- {\bf
a})^2 +a_0
\label{hamiltonian}
\end{eqnarray}
The velocity of
the center of the wave
packet is
\begin{eqnarray}
v^i &\equiv& {d\over dt}<\psi|x^i|\psi>={i\over\hbar} <\psi|
[H,x^i]|\psi>
\nonumber \\&& = {1\over m} <\psi| p^i-a^i |\psi>
\label{velocity}
\end{eqnarray}
The force is
\begin{equation}
m{dv^i\over dt} ={i\over\hbar} <\psi| [H, p^i-a^i]|\psi>
\label{nrforce}
\end{equation}
This gives the force to be the same as the
spatial components of (\ref{force}), in the present limit.

Suppose now that the dipole is made of
two particles with charges $q,-q$, masses $m_1,m_2$ and
coordinates
$ {\bf x}_1 , {\bf x}_2$. It can be shown that the oribital
angular momentum of the system about an arbitrary origin
\begin{equation}
{\bf x}_1 \times {\bf p}_1 + {\bf x}_2 \times {\bf p}_2
={\bf x}\times ({\bf
p}_1 + {\bf p}_2) + {\bf r}\times {\bf p}
\label{am}
\end{equation}
where ${\bf x}=(m_1+m_2)^{-1}(m_1 {\bf x}_1+ m_2 {\bf x}_2)$
is the center of
mass, ${\bf r}={\bf x}_1 - {\bf x}_2$ the relative
coordinates, and
\begin{equation}
{\bf p}={1\over m_1+m_2}(m_2 {\bf p}_1- m_1 {\bf p}_2).
\end{equation}
Then,
$[p_i,r_j]=-i \hbar
\delta_{ij}$ and $[p_i,x_j]=0$. The last term in (\ref{am}),
${\bf L}\equiv {\bf
r}\times {\bf p}$, is the total orbital angular momentum
about the center of
mass, and is invariant under Galilei boosts. It satisfies
$$[L_i,L_j]=i\hbar \epsilon_{ijk}L_k,[L_i ,r_j]= i\hbar
\epsilon_{ijk}r_k,
[L_i ,x_j]=0.$$

In general, the magnetic moment
\begin{equation}
{\bf \mu} = \gamma_L {\bf L} +\gamma_S {\bf S}
\end{equation}
where $\bf S$ is the total spin. The electric dipole moment
${\bf d}=q{\bf r}$.
It follows that
\begin{eqnarray}
[ \mu_i,\mu_j ]  &=&  i\hbar\epsilon_{ijk}( {\gamma_L}^2
L_k+{\gamma_S}^2
S_k ), ~[d_i,d_j]=0,\nonumber \\&&
[ \mu_i ,d_j ]  =  i\hbar \gamma_L
\epsilon_{ijk}d_k
\end{eqnarray}
In the rest frame of the dipole, i.e. a frame in which
(\ref{velocity}) is zero,
from (\ref{force}),  $f^0=0$ and
$f^i =<\psi|{G^{i}}_0|\psi>$. In the present low energy
limit, the latter force is
\begin{eqnarray}
{\bf f}&=& \nabla ({\bf \mu}\cdot
{\bf B}+ {\bf d}\cdot {\bf E})+ {1\over c}{\partial\over
\partial t}
({\bf d}\times {\bf B} - {\bf \mu}\times {\bf E})
\nonumber \\&& -
{\gamma_S^2\over c} ({\bf S}\times {\bf B})\times {\bf E}
-{\gamma_L^2 \over c}({\bf L}\times {\bf B})\times {\bf E}
\nonumber \\&&
-{\gamma_L\over c} ({\bf B}\times {\bf d})\times {\bf B}
-{\gamma_L \over c}({\bf E}\times {\bf d})\times {\bf E}
\label{leforce}
\end{eqnarray}
(\ref{leforce}) may also be obtained from (\ref{nrforce}).
$\bf \mu$, $\bf d$.
$\bf S$ and $\bf L$ in (\ref{leforce}), quantum
mechanically, are expectation
values of the corresponding operators. The force in an
arbitrary inertial frame
may be obtained by Lorentz transforming the above $f^\mu$ to
this frame.
Of the terms in (\ref{leforce}) that are non linear in the
electromagnetic
field, $-(\gamma_S^2/c) ({\bf S}\times {\bf B})\times {\bf E}$
was previously
obtained by the author \cite{an1989,an1989a} and may be
experimentally
detectable \cite{wa1997}. The remaining three non linear
terms are new.
Particularly interesting are the last two terms in
(\ref{leforce}), which are due to
the combined effect of the electric and magnetic dipole
moments.

An experiment that would detect a topological effect due to
$D_0$, which is
analogous to the scalar AB effect \cite{ah1959}, is the
following. Split a beam
of identical neutral particles with electric dipole moment
$\bf d$, but no
magnetic dipole moment, into two, and send one beam between
a pair of capacitor
plates that are initially neutral. The beam is so weak
that at most one
particle is inside the capacitor at any given time. The beam
is also polarized
so
that $\bf d$ is parallel to the electric field in the
capacitor. When each
particle is inside the capacitor turn on the homogeneous
electric field of the
capacitor and
turn it off before the particle leaves the capacitor plate.
Then the phase shift
due to $D_0$ is \cite{wi}
\begin{equation}
 - {1\over \hbar} \int_0^T D_0 dt= {1\over \hbar} \int_0^T
{\bf d}\cdot {\bf E} dt=(ETd/\hbar)
\end{equation}
This is analogous to
the experiment to
detect
the topological phase shift due to $M_0$ proposed
by Zeilinger \cite{ze1985} in analogy with the scalar AB effect
\cite{ah1959}, and by the author \cite{an1989a} as the scalar phase
shift corresponding to the vector phase shift considered by Aharonov and
Casher \cite{ah1984} in the interference of
magnetic moment around a
line charge.
This scalar phase shift has been
experimentally observed
\cite{ze1993,al1992}.

In the above mentioned three interferometry experiments
with electric or magnetic dipoles,
it is  the
expectation value of the quantum force
operator, or the classical force, that vanishes, whereas the
quantum force,
$m(d^2\hat x^i/dt^2)$, is non zero \cite{an1994}. For
example, in the second of
the above three experiments \cite{ze1985,an1989a}, the
linear terms in the
quantum force vanish because of the homogeneity of the
magnetic field, on using
a Maxwell's equation. But the non linear term $-\gamma_S^2
(\hat{\bf S}\times
{\bf B})\times {\bf E}$ does not vanish. It is the
expectation value of the
latter term, or the corresponding classical force term, that
vanishes. All three
experiments are weakly topological because, along the
interfering beams,
(\ref{weak}) holds whereas (\ref{strong}) does not hold.
This is analogous to
the Stern-Gerlach experiment with the spin perpendicular to
the inhomogeneity
of the magnetic field for which also (\ref{weak}) holds
whereas (\ref{strong})
does not hold. In the latter experiment the existence of the
quantum  force is
obvious because it splits the beam into two, even though the
classical force is
zero. On the other hand, the AB effect \cite{ah1959} is
strongly topological, as
defined above, because $G_{\mu\nu}=e F_{\mu\nu}=0$
in this case.

However, in the following proposed interferometric experiments
(\ref{strong}) holds, and hence
the quantum force is zero and the phase shifts are strongly
topological and non
local. Subject the entire interferometer to a homogeneous
electric or magnetic
field. Along each beam the magnetic or electric dipole
moment is rotated by
$180^0$ and is rotated back again by $180^0$ by suitable
oscillating fields
after the beams have traveled distances $L_1$ and $L_2$,
respectively. The
contributions to the phase
shift due to the dipole flips
on one beam is
canceled by the ones in the other beam by doing them
identically. By varying $L_1-L_2$ on which the phase shift depends, it
can be shown clearly that the phase shift is due to the quantum force
free interaction of the dipole. In a neutron interference experiment,
for example, if the entire interferometer subject to a homogeneous
electric field $\bf E$, the phase shifts for neutron spin components
parallel and anti-parallel to ${\bf E}\times {\bf v}$ are \cite{an1982}
\begin{equation}
\pm {2|{\bf M}|\over \hbar} ( L_1-L_2) =\pm {2\mu E\over \hbar} (
L_1-L_2) .
\end{equation}

The most sensitive method for detecting the electric dipole
interaction is by
the electric resonance method described by Ramsey
\cite{ra1956}. Here a
polarized molecular beam is deflected and refocused by two
inhomogeneous
electric fields. In between, a homogeneous and oscillatory
electric fields are
applied. The fields are chosen so that an allowed electric
dipole transition
occurs due to the latter fields which makes the intensity of
the refocused beam
maximum. Introduce a magnetic field ${\bf B}$ in
the intermediate
region so that ${\bf v}\times {\bf B}$ is in the direction
of the homogeneous
electric field. This shifts the intensity from the maximum
which may easily be
detected.

I thank Kuniharu Kubodera and Norman F. Ramsey for useful discussions.


\begin{thebibliography}{99}
\bibitem{dual}
The dual electromagnetic field $ {F*}_{\mu\nu} \equiv {1\over 2}
\epsilon_{\mu\nu\rho\sigma}
F^{\rho\sigma}$ where
$\epsilon_{\mu\nu\rho\sigma}$ is the volume form that is the
completely anti-
symmetric tensor with $\epsilon_{0123}=1$ in every inertial
frame.
\end{thebibliography}
\end{document}